\documentclass[draft]{agujournal}

\usepackage{amsmath,amssymb,amsthm}
\usepackage{caption}
\usepackage{comment}
\usepackage{placeins}
\usepackage{array}

\journalname{Space Weather}

\begin{document}

%
%


\title{Improving Solar Wind Forecasting using Data Assimilation}

%
%




\authors{Matthew Lang \affil{1}, Jake Witherington \affil{1}, Harriet Turner \affil{1}, Mathew J. Owens \affil{1}, Pete Riley \affil{2}}


\affiliation{1}{Department of Meteorology, University of Reading, Reading, Berkshire, UK}
\affiliation{2}{Predictive Science Inc., 9990 Mesa Rim Rd., Suite 170, San Diego, CA, 92121, USA}





\correspondingauthor{Matthew Lang}{matthew.lang@reading.ac.uk}




\begin{keypoints}
\item Data assimilation (DA) is shown to improve solar wind speed forecasts over the interval 2007-2014.
\item Initialising forecasts with DA leads to a $31.4\%$ reduction in solar wind speed root mean-square error (RMSE) compared to forecasts with no DA.
\item DA can remove significant positive correlations between forecast RMSE and latitude of STEREO-B
\end{keypoints}

%
%


\begin{abstract}
Data Assimilation (DA) has enabled huge improvements in the skill of terrestrial operational weather forecasting. In this study, we use a variational DA scheme with a computationally efficient solar wind model and in situ observations from STEREO-A, STEREO-B and ACE. This scheme enables solar-wind observations far from the Sun, such as at 1 AU, to update and improve the inner boundary conditions of the solar wind model (at $30$ solar radii). In this way, observational information can be used to improve estimates of the near-Earth solar wind, even when the observations are not directly downstream of the Earth. This allows improved initial conditions of the solar wind to be passed into forecasting models. To this effect, we employ the HUXt solar wind model to produce 27-day forecasts of the solar wind during the operational lifetime of STEREO-B (01 November 2007 - 30 September 2014). In near-Earth space, we compare the accuracy of these DA forecasts with both non-DA forecasts and simple corotation of STEREO-B observations. We find that $27$-day root mean-square error (RMSE) for STEREO-B corotation and DA forecasts are comparable and both are significantly lower than non-DA forecasts. However, the DA forecast is shown to improve solar wind forecasts when STEREO-B's latitude is offset from Earth, which is an issue for corotation forecasts. And the DA scheme enables the representation of the solar wind in the whole model domain between the Sun and the Earth to be improved, which will enable improved forecasting of CME arrival time and speed.

\end{abstract}

\section{Introduction}
\FloatBarrier
\begin{center}
	\begin{figure}
		\begin{center}
			\includegraphics[width=\textwidth]{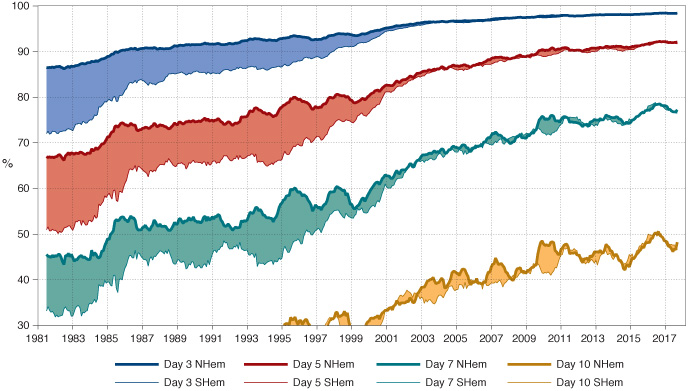}
			\caption{The 12-month running mean of the anamoly correlation for ECMWF's (European Centre for Medium-range Weather Forecasting) $500~\mathrm{hPa}$ geopotential height forecasts in the Earth's northern and the southern hemisphere at $3-$day (blue), $5-$day (red), $7-$day (green) and $10-$day (yellow) lead-times. The shaded regions show the difference between the northern (top) and southern (bottom) hemispheres, which is largely a result of differing observational coverage. Image reproduced from \url{https://www.ecmwf.int/en/about/media-centre/news/2018/noaa-satellite-launch-20-years-ago-marked-start-new-era}}
			\label{fig:500geopotECMWF}
		\end{center}
	\end{figure}
\end{center}

The solar wind is a continuous outflow of plasma and magnetic flux which fills the heliosphere (e.g., \citet{owens2013heliospheric}). Variations in the solar wind properties in near-Earth space can lead to adverse effects \citep{hapgood2011}, such as disruption to power and communications systems, and health hazards for astronauts and aircrew on flights over the poles. As a result, it is vital to accurately and efficiently forecast solar-wind conditions in near-Earth space. However, forecasting ahead by more than about one hour (the nominal travel time to Earth for solar wind from L1, the gravitational null 0.01AU ahead of Earth on the Earth-Sun line, where observing spacecraft are routinely positioned) requires  prediction of the solar wind conditions near the Sun. This is typically achieved using empirical relations to the coronal magnetic field \citep{arge2003Improved, mcgregor2011, riley2015}, which is in turn determined using extrapolation from the observed photospheric magnetic field \citep{mackay2012, linker1999}. These empirical models approximate the solar wind conditions at $\sim 21-30$ solar radii ($r_S$). From here, the solar wind is propagated to Earth, typically using a numerical magnetohydrodynamic (MHD) model, such as ENLIL \citep{odstrcil2003modeling}, HelioMAS \citep{riley2012corotating}, HelioLFM \citep{Merkin2016MHD} or EUHFORIA \citep{poedts2017euhforia}, with no further observational constraints.

In this study, however, we apply a further step using Data Assimilation (DA), by assimilating available in-situ observations of the solar wind, before forecasting the solar wind. Data assimilation is the process by which observational data is incorporated into numerical models to improve knowledge of the trajectory, and uncertainties, of the model state. In operational meteorological
models, data assimilation is commonly used to generate an optimal initial state to be used in subsequent forecasts through the use of prior knowledge of the state (e.g., a previous forecast) and observational data obtained from satellites/weather stations, etc. Data assimilation is considered a vital step in meteorological forecasting, reducing the impact of the `butterfly effect' (where small changes in initial conditions lead to large differences in model evolution), and hence improving the forecast skill. Improvements in numerical weather predictions have gone hand-in-hand with improvements in the implementation of the observation network and data assimilation methods into the models. This has led to huge improvements in the lead-times over which forecasts are accurate  \citep{kalnay2003atmospheric}. This can clearly be seen in Figure \ref{fig:500geopotECMWF}.  In the late nineties, a significant improvement was made to data assimilation methodology and satellite observations were included in the data assimilation, vastly increasing observational coverage. The assimilation of satellite observations led to vast improvements in Southern Hemisphere forecasts especially, leading to the closing of the gap in Northern and Southern Hemisphere forecast skill. Space weather forecasting has not yet exploited the full potential available from implementing data assimilation methods in forecasting models.

Previous DA work in space-weather forecasting has focused mostly on the photosphere, such as the Air Force Data Assimilative Photospheric Flux Transport (ADAPT) model \citep{arge2010air}, or the ionosphere (\citep{bust2008Ion, durazo2017local}. Solar wind data assimilation is less mature and much work is still required in this area. \citet{lang2017DA} connected the EMPIRE (Employing MPI for Researching Ensembles) \citep{browne2015simple} data assimilation library to the ENLIL solar wind model in the first implementation of complex DA methodologies into a solar wind model. This work highlighted the sharp differences between terrestrial and solar wind DA, specifically that the solar wind is highly driven by the Sun with relatively little chaotic behaviour. Data assimilation updates were swept away if they did not (permanently) update the inner boundary and these updates did not have any influence at Earth unless the observation was downstream of Earth. This means that any future DA methodologies need to be developed to reflect this and allow DA updates to persist in the system and influence near-Earth space. Aiming to update the inner boundary conditions, \citet{lang2019var} generated a variational data assimilation scheme with a simple solar wind propagation model (see \citet{riley2012corotating} and \citet{owens2017probabilistic} for more details of the solar wind model) that could map information from in-situ observations at near-Earth orbit to the inner solar wind model boundary. This was shown to give improved and persistent updates to the solar wind over a $27$-day period. However, this demonstration of improved solar wind reconstruction was not directly analogous to how the method would work in a forecast situation. 

In this study, we will use the data assimilation methodology developed in \citet{lang2019var}, referred to here as the Burger Radial Variational Data Assimilation (BRaVDA) scheme, to investigate how data assimilation of in-situ observations affects solar wind forecast (strictly speaking, hindcast) skill. The BRaVDA scheme will be used to assimilate observations from the twin Solar-Terrestrial Relations Observatory (STEREO) \citep{kaiser2008stereo} spacecraft, in Earth-like orbits about the Sun, and the Advanced Composition Explorer (ACE) \citep{stone1998advanced} in near-Earth space. The BRaVDA scheme will be used to generate $27-$day reanalysis sets over the operational lifetime of the STEREO-B satellite (01 November 2007 - 30 September 2014). For each $27$-day DA analysis, a $27-$day solar wind forecast will be produced using the HUXt (Heliospheric Upwind Extrapolation with time dependence) solar wind forecasting model \citep{owens2020huxt}, initialised with the results of the BRaVDA scheme. These forecasts will be presented and their accuracy will be verified against in-situ observations at STEREO and ACE. As a benchmark, a simple corotation forecast from the STEREO-B spacecraft (which is behind Earth in its orbit for the time-period of interest) will also be produced at L1 for comparison with ACE observations. To produce this corotation forecast, the solar wind structures will be assumed to be steady and to statically rotate with the Sun. Therefore, the observations at STEREO-B are predicted to be the solar wind conditions at L1 $x$ time later, where $x$ is the time it takes for the Sun to rotate between the longitudes of STEREO-B and L1 \citep{thomas2018corot, kohutova2016Improving}.

This paper will describe the BRaVDA scheme and the HUXt forecasting model. Numerical experiments will then be presented, analysing the seven-year BRaVDA reanalyses and subsequent HUXt forecasts. Statistical analysis shall be performed and the improvements due to data assimilation will be discussed. In the final section, we shall make conclusions and discuss future developments.

\section{The BRaVDA scheme}
\label{sec:bravda}

\begin{figure}
    \centering
    \includegraphics[width=\textwidth]{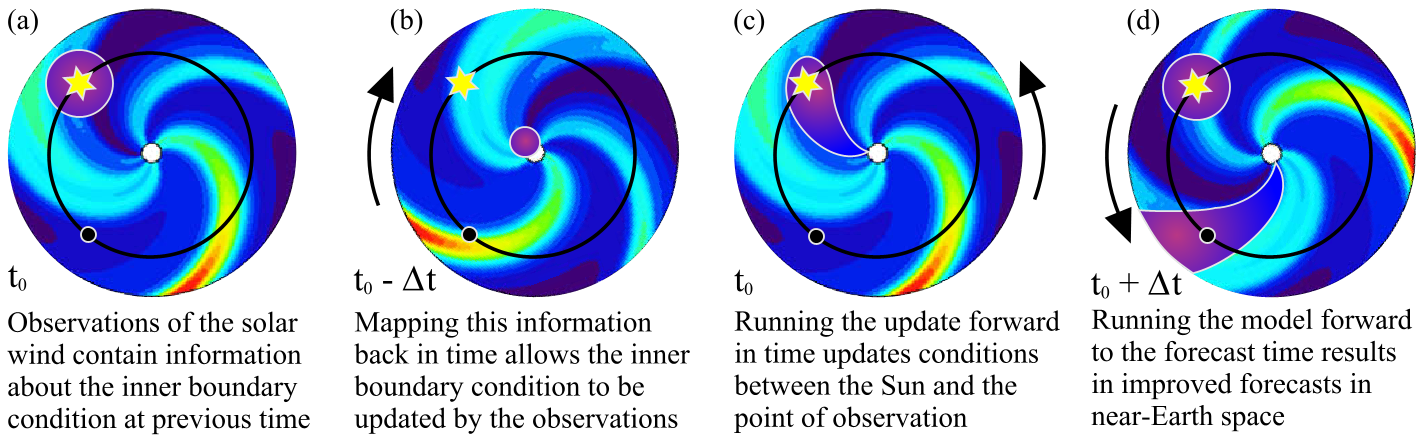}
    \caption{A schematic of how the BRaVDA scheme works from \citep{lang2019var}. The inner boundary (the white circle) is updated using information from observations from a position behind Earth in its orbit (the yellow star). The updated model conditions (the purple regions) persist in the model domain such that they impact forecasts at Earth’s location (the black circle). Updating the inner boundary stops updates from being advected out of the model domain by the supersonic radial solar wind flow. (a) Observations of the solar wind contain information about the inner-boundary condition at previous time. (b) Mapping this information back in time allows the inner boundary to be updated by the observations. (c) Running the update forward in time updates conditions between the Sun and the point of observation. (d) Running the model forward to the forecast time results in updated forecasts in near-Earth space.}
    \label{fig:BRaVDAoverview}
\end{figure}

The BRaVDA scheme is a variational data assimilation scheme \citep{lang2019var}, to map information from in-situ observations, typically near 1 AU, back to a solar wind model's inner boundary (in this case, $30$ solar radii [$r_S$]). Figure \ref{fig:BRaVDAoverview} outlines qualitatively how the BRaVDA scheme works: our current estimate for the solar wind speed at the inner boundary is mapped out to the observation radius using our solar wind propagation model. Information from the observation (including its uncertainty) is then mapped back towards the Sun (Figure \ref{fig:BRaVDAoverview}a-\ref{fig:BRaVDAoverview}b), where it is used to update the inner boundary of the solar wind model. The updated inner boundary is propagated back out using the solar wind model (Figure \ref{fig:BRaVDAoverview}c), the results of which can then be used to produce forecasts of the solar wind speed at Earth (Figure \ref{fig:BRaVDAoverview}d).

The solar wind model used by BRaVDA is based upon a solar wind speed propagation model developed by \citet{riley2011mapping}, which maps the (two-dimensional) equatorial solar wind speed over the heliocentric domain from $30 r_S$ to $236 r_S$ from the Sun. Given the 2D nature of this model, we only represent a single heliolatitude. In reality there is typically a latitudinal offset between observing spacecraft and the desired forecast location. For example, the STEREO spacecraft are in the ecliptic plane, which is inclined by $7.25^\circ$ to the heliographic equator. This introduces a representivity error (an error that arises as a result of representing an observation incorrectly in a model; other examples include an observation moved to the nearest gridpoint or a parameterisation estimating a sub-gridscale process) into the system. A solar wind DA scheme in a full 3-dimensional solar wind model could relax this assumption, though the problem of determining the latitudinal representivity error of the observations remains.

In data assimilation, we aim to reconstruct the true state vector, $\mathbf{x^t}$, an $N_x$-dimensional vector that contains the values of the quantities of interest at all gridpoints, using prior knowledge about the system (e.g., a previous forecast), $\mathbf{x^b}$, and observations, $\mathbf{y}$. The variables within a state vector for meteorological applications can include the precipitation over Africa, the mean sea level pressure over France, wind speed/direction in Llanfairpwllgwyngyllgogerychwyrndrobwllllantysiliogogogoch, etc. depending upon the purpose of the numerical model. In order to do this, the errors in the prior and observational information is typically assumed to be normally distributed and the numerical model is assumed to be linear. This allows a cost function to be defined, based upon the relative uncertainties in the prior errors, $\mathbf{B}$, and the observation errors, $\mathbf{R}$. Through construction, the state that minimises the cost function maximises the posterior distribution (the probability distribution of the state given the observations, $p(\mathbf{x}|\mathbf{y})$, which is proportional to the likelihood distribution, $p(\mathbf{y}|\mathbf{x})$, multiplied by the prior distribution, $p(x)$, through Bayes' Theorem \citep{bayesTheorem1763}), hence giving us the mode of the posterior distribution. 

In the BRaVDA system, the state vector that we wish to estimate is the solar wind speed at the inner-boundary at all longitudinal points (i.e., a $128$-dimension vector, $\mathbf{v_0}$). The prior error covariance matrix, $\mathbf{B}$, is estimated by an ensemble of inner boundary solar wind speeds generated using the MAS (Magnetohydrodynamics Around a Sphere \citep{linker1999}) model (further details about how this ensemble is created can be found at \citet{owens2017probabilistic}) as:
\begin{equation}
\label{eq:priErrCovEns}
\mathbf{B} \approx \frac{1}{M-1} \sum_{m=1}^{M}\left[\left(\mathbf{v^{(m)}_0}-\overline{\mathbf{v^M_0}}\right)\left(\mathbf{v^{(m)}_0}-\overline{\mathbf{v^M_0}}\right)^T\right],
\end{equation}
where $\overline{\mathbf{v^M_0}}=\frac{1}{M}\sum_{m=1}^{M}\left[v^{(m)}_0\right]$ is the MAS ensemble mean and $M=576$ is the number of ensemble members. The prior state, $\mathbf{v^b}$, is a solar wind speed randomly drawn from the normal distribution, $\mathcal{N}(\overline{\mathbf{v^M_0}},\mathbf{B})$, defined by the probability distribution function:
\begin{equation}
    p(x) = \left(2 \pi\right)^{\frac{-N_x}{2}} \det\left(\mathbf{B}\right)^{\frac{-1}{2}} \exp\left(-\frac{1}{2}\left(x - \overline{\mathbf{v^M_0}}\right)^T B^{-1} \left(x - \overline{\mathbf{v^M_0}}\right) \right),
\end{equation}
where $\det\left(\mathbf{B}\right)$ is the determinant of $\mathbf{B}$.

The BRaVDA scheme also utilises the Strong Constraint \citep{howes2017strong}, which makes an additional assumption that the numerical model is perfect (i.e., the numerical model contains no errors). This typically produces a poorer result than a system that allows for model error (the Weak Constraint), however incorporating model error is a complex operation \citep{parEstEvensen1998,lang2016ParsationEst}, and is beyond the scope of the present study. The BRaVDA Cost Function ($\mathcal{J}\left(\mathbf{v_0}\right)$, a full derivation of which can be found in Appendix A of \citet{lang2019var}), that is to be minimised, is given by:
\begin{equation}
\label{eq:costFunc}
\mathcal{J}\left(\mathbf{v_0}\right)=\frac{1}{2}\left(\mathbf{v_0}-\mathbf{v^b}\right)^T\mathbf{B}^{-1}\left(\mathbf{v_0}-\mathbf{v^b}\right)+\frac{1}{2}\sum_{k=0}^{N_y}\left[\left(\mathbf{y_k}-\mathcal{H}_k(\mathbf{v_0})\right)\mathbf{R_k}^{-1}\left(\mathbf{y_k}-\mathcal{H}_k(\mathbf{v_0})\right)\right],
\end{equation}
where $\mathbf{y_k}$ is the $k^{th}$ observation, $N_y$ is the number of observations, and $\mathcal{H}_k$ is the observation operator that is a function that maps the state to the observation space (i.e. $\mathcal{H}_k \left(\mathbf{x_0}\right)$ tells us what the numerical model `thinks' the observation should be). The observations used in this study are in-situ observations from spacecraft such as STEREO and ACE, but can also be expanded to include remotely sensed observations such as Interplanetary Scintillation (IPS; e.g., \citet{breen2006}) or Heliospheric Imagers (HI; e.g., \citet{eyles2009}), albeit these require a far more complex observation operator than is currently utilised for in-situ observations in BRaVDA. $\mathbf{R_k}$ is the observation error covariance matrix and approximating it fully is a large research topic in its own right \citep{owens2020latRep}, as it will contain representivity errors. The BRaVDA scheme currently only supports diagonal $\mathbf{R}$ matrices that are either constant or proportional to the prior state, which implies that all the observations are assumed to be independent of each another and with equal uncertainty leading to potential inaccuracies (the addressing of which is beyond this study).

The $\mathbf{v_0}$ that minimises the cost function is referred to as the analysis or posterior state and is given the notation $\mathbf{v^a}$ in this manuscript. Obtaining $\mathbf{v^a}$ is nontrivial and can be done by using a finite differences approach or by evaluating the gradient of the cost function directly \citep{bannister2007Elementary}, however, for higher-dimensional systems, these methods are not practical. A more efficient method of calculating the gradient is to use the adjoint method, a method of sensitivity analysis that efficiently computes the gradient of a function \citep{errico1997adjoint}. The adjoint equations for the BRaVDA scheme are derived explicitly in \citet{lang2019var}. The BRaVDA scheme is capable of producing a $27$-day analysis assimilating STEREO-A, B and ACE data over the period ($\sim 128$ obs each) in $6~\mathrm{min}~36s \pm 6.1s$ on a home laptop with $8$GB of RAM and an Intel Core i5-7300HQ 2.5GHz processor. Thus the scheme is suitable for near-real time forecasting.

This adjoint method is extremely powerful once computed and its implementation enabled the great improvements in forecast skill seen by ECMWF in the $1990s$ (see Figure \ref{fig:500geopotECMWF}). However, the adjoint equations are unique for each model, which means that they can be difficult to produce for complex models.  Future studies will use DA methodologies that remove the need to use an adjoint and may allow for more nonlinear behaviour, such as 4DEnsVar \citep{chumps2017Part1,chumps2017Part2}, or a smoother-based ensemble DA method, such as the Iterative Ensemble Kalman Smoother \citep{bocquet2014IEnKS}.

\section{The HUXt solar wind model}

\label{sec:huxt}

The HUXt (Heliospheric Upwind eXtrapolation with time dependence) solar-wind model is a reduced-physics solar-wind model based upon Burgers' inviscid equation \citep{owens2020huxt}. In order to make the solar-wind model as computationally efficient as possible, a number of physical assumptions and approximations are required, beyond those typically used in three-dimensional magnetohydrodynamics. These are that in the heliosphere, pressure gradient and gravitation terms are small compared with the flow momentum and hence can be neglected. A further simplification is that we only consider variations in the radial direction. See also \citet{riley2011mapping}. 

Using 40+ years of hindcasts of ambient solar wind, the HUXt model was shown to emulate the full three-dimensional MHD solution using the same $30r_S$ boundary conditions, with solar-wind speeds throughout the simulation domain agreeing to within $6\%$ \citep{owens2020huxt}. The biggest difference between the two solutions is that HUXt produces sharper boundaries between slow and fast wind than the MHD solutions of HelioMAS and ENLIL (see \citep{owens2020huxt} for more details).

\section{Experimental setup}
\label{sec:expsetup}
\begin{figure}
	
	\includegraphics[width=\textwidth]{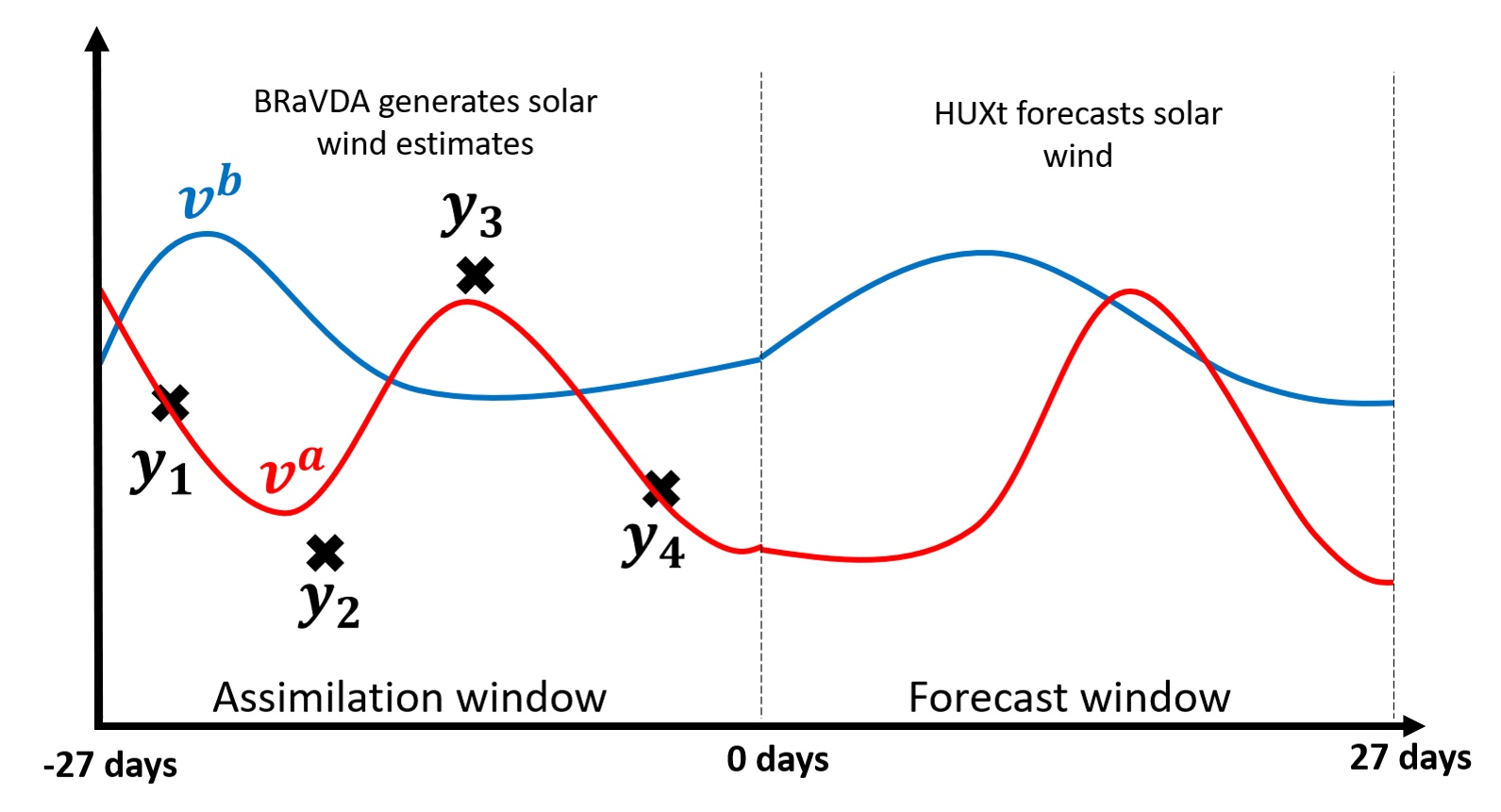}
	\caption{Schematic showing how BRaVDA and HUXt are used to generate solar-wind estimates and forecasts. The prior solar wind ($\mathbf{v^b}$, blue), is generated using the radial solar-wind propagation scheme defined from \citet{riley2011mapping}. The BRaVDA scheme uses Equation \eqref{eq:costFunc} and the observations ($\mathbf{y_k}$, black crosses), to generate an analysis ($\mathbf{v^a}$, red), for the $27$ days prior to the forecast start-time (the assimilation window). $\mathbf{v^b}$ and $\mathbf{v^a}$ are used to initialise HUXt forecasts of the solar-wind speed for the next $27$ days (forecast window).}
	\label{fig:fcSchem}
\end{figure}

For each consequent $27$-day period from 01 January 2007 - 02 November 2014, the MAS ensemble (available from http://www.predsci.com/mhdweb/home.php) is constructed for that period and the prior state, $\mathbf{v^b}$, and prior error covariance matrix, $\mathbf{B}$, are approximated as described in Section \ref{sec:bravda}. 

The observation error covariance matrix, $\mathbf{R}$, is also generated for each $27$-day window. As in \citet{lang2019var}, we specify the observation error covariance matrix as a diagonal matrix with standard deviation approximated as $10 \%$ of the mean prior solar-wind speed at the observation radius. This is a somewhat arbitrary value, as the true $\mathbf{R}$ matrix will include a term that varies depending upon the observation's latitudinal offset from Earth and phase of solar cycle, etc., but it provides a starting point from which we can progress and test such assumptions. This gives:
\begin{align}
\label{eq:R01xMean}
R_{kk}&=(0.1\overline{\mathbf{v^b_{r_k}}})^{2} \notag\\
&=(0.1 \sum_{i=1}^{N_{\phi}}\left[v^b_{r}(\phi_i)\right])^{2},
\end{align}
where $0 \leq k \leq N_y$ is the observation number and $N_{\phi}=128$ is the number of longitude points.

The BRaVDA scheme is run with these parameters for consecutive $27$-day windows (each approximating $1$ solar rotation from Earth's perspective) spanning the seven-year operational window of STEREO-B, assimilating all available observations from STEREO-A, STEREO-B and ACE. This produces initial conditions for forecasting the next $27$-day period. The HUXt model is used to produce the forecast for the next $27$-day window, with no additional observational constraints. This is summarised in Figure \ref{fig:fcSchem}.

\section{Results}
 \label{sec:results}
 
\subsection{Forecasting results over the whole $7-$year period}


\begin{figure}
	\begin{center}
		\includegraphics[width=\textwidth]{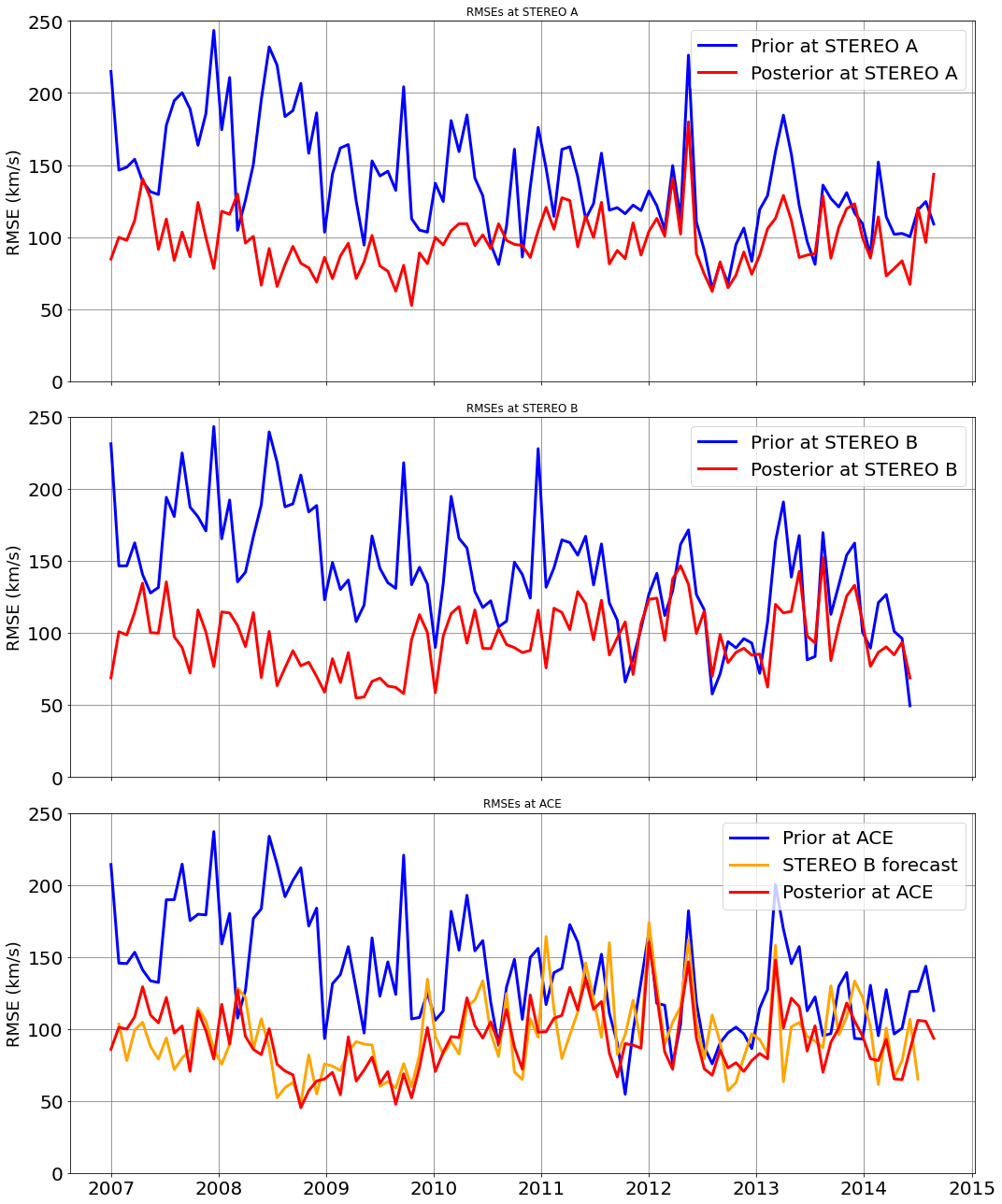}
		\caption{Root mean-square error (RMSE) of the solar wind speed averaged over each $27$ day window. The top row gives the RMSEs at STEREO-A, the middle row gives the RMSEs at STEREO-B and the bottom row gives the RMSEs at ACE. The blue lines indicate the prior (no DA) forecast RMSEs, the red lines indicate the posterior (including DA) forecast RMSEs and the orange line on the bottom plot gives the STEREO-B corotation forecast at ACE.}
		\label{fig:7year27dRMSE}
	\end{center}
\end{figure}

\begin{figure}
	\begin{center}
		\includegraphics[width=\textwidth]{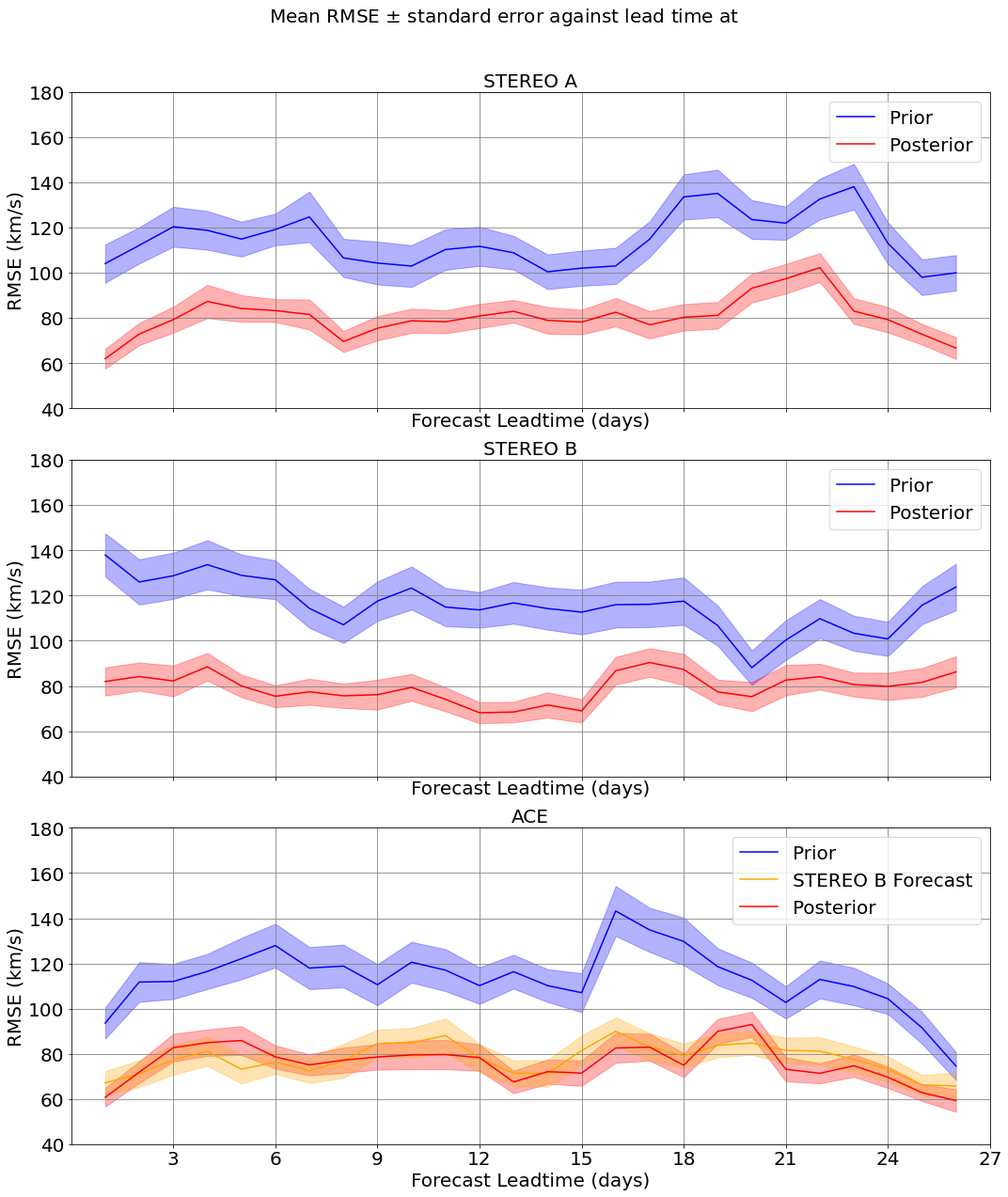}
		\caption{The average forecast Root Mean Squared Errors (RMSEs) of the solar-wind speed plotted against forecast lead-time, for the whole 2007-2014 interval. The top row corresponds to the RMSEs at STEREO-A, the middle row corresponds to the RMSEs at STEREO-B and the bottom row corresponds the RMSEs at ACE. The blue lines indicate the prior (no DA) forecast RMSEs, the solid red lines indicate the posterior (DA) forecast RMSEs and the solid orange line on the bottom plot gives the STEREO-B corotation forecast at ACE. The solid lines indicate mean RMSEs while the shaded regions span one standard error.}
		\label{fig:mean27dRMSEAll}
	\end{center}
\end{figure}

\begin{table}
	\begin{center}
		\begin{tabular}{ | >{\centering\arraybackslash}m{1.9cm} | >{\centering\arraybackslash}m{2.5cm} | >{\centering\arraybackslash}m{2.5cm} | >{\centering\arraybackslash}m{3.6cm} | }
			\hline
			 Observation location & Prior RMSE (km/s) & Posterior RMSE (km/s) & STEREO-B corotation RMSE (km/s) \\ \hline
            STEREO-A & $114.507 \pm 8.618$ & $80.419 \pm 5.626$ &  N/A \\ \hline
            STEREO-B & $115.965 \pm 9.111$ & $79.384 \pm 5.870$ & N/A \\ \hline
            ACE & $113.349 \pm 8.416$ & $76.089 \pm 5.387$ & $77.752 \pm 5.951$ \\ \hline \hline
            All avg. & $114.607 \pm 8.715$ & $78.631 \pm 5.628$ & N/A \\
			\hline
		\end{tabular}
	\end{center}
	\caption{Table showing mean Root Mean Squared Errors (RMSEs) $\pm$ their standard errors for the prior, posterior and STEREO-B corotation forecasts at each observation location over the whole $7$ year window.}
	\label{tab:meanFc7year}
\end{table}

\begin{figure}
	\begin{center}
		\includegraphics[width=\textwidth]{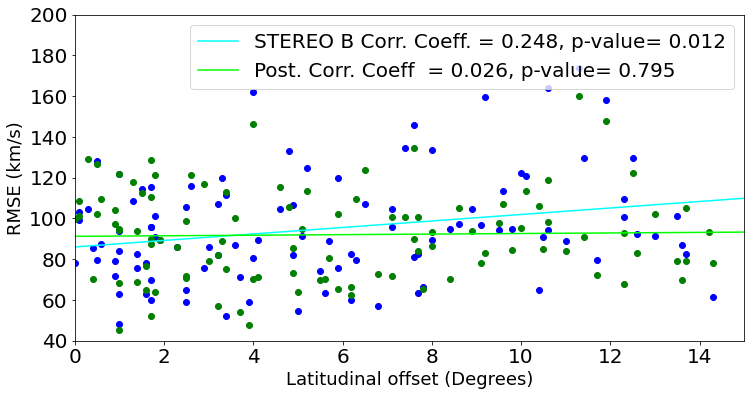}
		\caption{The 27-day forecast Root Mean Squared Errors (RMSEs) of the solar-wind speed plotted against the absolute difference between STEREO-B and ACE's latitude, for the entire seven-year period. The blue points correspond to the STEREO-B corotation forecast and the green points relate to the DA posterior forecast. The blue and green lines represent the line of best-fit for the corotation and posterior forecasts, respectively. The legend gives the Pearson's product-moment correlation coefficient indicating the strength of the correlation between the data and the line of best-fit, along with its associated p-value. The STEREO-B corotation forecast RMSE has a correlation coefficient of $0.248$ with STEREO-B latitude, which is significant at the level of $p = 0.012$.}
		\label{fig:27dlatAll}
	\end{center}
\end{figure}

This section describes the results of the forecasts produced using the data assimilation over the whole seven-year window. As mentioned in the previous sections, $27$-day forecasts of the solar wind are produced for the prior state (generated from an ensemble of MAS runs) and the posterior state (the resultant solar wind from the DA). At ACE, we also compared against the STEREO-B corotation forecast. This involves time-shifting STEREO-B observations by $\Delta t$, given by:
\begin{equation}
\label{eq:corotTime}
\Delta t = \frac{\Delta \phi_{B}}{\Omega_{ROT}},
\end{equation}
where $\Delta \phi_{B}$ is the longitude difference between STEREO-B and Earth. See \citet{thomas2018corot, kohutova2016Improving} for more details.

For each $27$-day forecast window, Root Mean Squared Error (RMSE) is computed over the whole $27$-day forecast window, i.e.:
\begin{equation}
\label{eq:RMSE7year}
RMSE_{27,y}=\sqrt{\frac{1}{N_{y}}\sum_{i=1}^{N_{y}}\left[\left(y_i-x_i\right)^{2}\right]},
\end{equation}
where $N_y$ corresponds to the number of STEREO-A, STEREO-B and ACE observations during the $27$-day forecast window, $y_i$ correspond to the STEREO-A, STEREO-B or ACE observations, and the $x_i$ are the prior/posterior state, at the relevant satellite's time/location.

This measures the accuracy of the solar wind forecasts over the whole 27-day forecast window. RMSE is a simple point-by-point analysis, which can be problematic for distinguishing between forecasts with insufficient variability and forecasts with small timing errors in events \citep{owens2018rmse}. However, in this particular application, it nevertheless provides a useful initial diagnostic of performance. More detailed forecast validation will be the subject of future research.

Figure \ref{fig:7year27dRMSE} shows the $27$-day forecast RMSEs against time. The posterior forecast RMSEs are lower than the prior forecasts RMSEs for the vast majority of windows, at all spacecraft locations. To be more specific, for the $105$ individual $27$-day forecasts generated: the posterior RMSEs are lower than the prior's for $95$ forecasts at STEREO-A, $92$ forecasts at STEREO-B and $97$ forecasts at ACE. This result indicates that the improvements made by the DA in the assimilation window are carried into the forecast window by the HUXt model.

In the first half of the seven-year period (i.e., before 2011), we note that the prior state has relatively high RMSEs. This is likely due to the Sun being in the solar-minimum phase, meaning that greater representivity errors due to latitudinal offsets are observed (see \citet{owens2020latRep} for more details). Also the solar wind speed is generally higher and more (spatially) variable at this time \citep{Owens2020Structure}. We can see that the data assimilation is able to produce more accurate solar wind speed forecasts at all observation locations for the vast majority of this time period.

In the latter half of the seven-year period, the solar-maximum phase implies the prior state has lower RMSEs due to reduced latitudinal representivity errors and lower solar-wind speed/variability \citep{owens2020latRep}. In addition, solar-maximum is associated with more transient events. These facts lead to a reduction in the improvement in DA forecast RMSEs from the posterior forecast (as the HUXt forecasts here assume the Sun is in steady-state). Nonetheless, the DA still leads to improvements in the forecast solar-wind speeds in comparison to the prior forecasts.

In the centre of the seven-year period (end of $2010$/start of $2011$), there appears to be regions where the prior, posterior and STEREO-B corotation forecasts produce very similar RMSEs. This is most likely due to the start of solar-maximum and STEREO-B being far from Earth (in both longitude and latitude). This corresponds to errors from the longitudinal offsets reducing as we move into solar-maximum, but corotation lead time and increased transients/solar-wind complexity compensating for these effects. These effects are difficult to disentangle and are currently being investigated to form part of a future study.

At ACE, we also compare the posterior forecast to the corotation forecast from STEREO-B. The two forecasts are comparable when looking at the $27$-day forecast RMSE. This is to be expected somewhat, as the data assimilation analysis incorporates the observation at STEREO-B within it. Nonetheless, the posterior forecast at ACE has lower $27$-day RMSEs for $63$ out of $105$ windows (that is, the posterior forecasts are better for $60 \%$ of windows). Note that the STEREO-B forecast does not exist for three windows, due to insufficient observations (but the posterior forecast still exists, as it can be based on ACE and STEREO-A data).

Figure \ref{fig:mean27dRMSEAll} shows the mean forecast RMSEs as a function of forecast lead-time at STEREO-A, STEREO-B and ACE. The mean RMSE for each forecast lead-time (in days) is calculated as:
\begin{equation}
\label{eq:RMSEleadTime}
RMSE_{t,y}=\sqrt{\frac{1}{N_w}\frac{1}{N_t}\sum_{l=1}^{N_w}\sum_{k=s_t}^{e_t}\left[\left(y_{l,k}-x_{l,k}\right)^{2}\right]},
\end{equation}
where $N_w$ is the number of $27$-day windows ($105$ in total), $N_t=e_t-s_t$ is the number of observations during the $t^{th}$ day, $y_{l,k}$ is the $k^{th}$ observation in the $l^{th}$ $27-$day window and $x_{l,k}$ is the state corresponding to $y_{l,k}$.

Looking at all observation locations in Figure \ref{fig:mean27dRMSEAll}, the posterior solar wind speed forecast has lower mean RMSE than the prior solar wind speed forecast for all lead-times by about $30-50 \mathrm{km/s}$. Table \ref{tab:meanFc7year} summarises this, with the mean RMSE reduction from prior to posterior being $35.976~\mathrm{km/s}$ over all observation locations (equivalent to a $31.4\%$ reduction in forecast RMSE over all observations). It is also the case that the standard errors for the posterior and STEREO-B corotation forecasts tend to be lower than that of the prior. This is further emphasised by the mean standard errors shown in Table \ref{tab:meanFc7year} that indicate that the posterior and STEREO-B corotation have standard errors of $5.628~\mathrm{km/s}$ and $5.951~\mathrm{km/s}$ respectively, whereas the prior forecast has standard errors averaging $8.715~\mathrm{km/s}$. This indicates the accuracy of the DA and corotation forecasts is more consistent than the prior. At ACE, the STEREO-B corotation forecast RMSE is very similar to the posterior forecast RMSE over all lead-times. However, the posterior forecast does have a slightly lower mean RMSE (see Table \ref{tab:meanFc7year}) and standard error than the STEREO-B forecast. Furthermore, the mean posterior forecast RMSE is lower for $16$ lead-times out of $27$ considered ($60\%$).

In Figure \ref{fig:27dlatAll}, the distribution of RMSEs for the forecasts is shown as a function of the absolute heliographic latitude difference between STEREO-B and ACE/Earth. The STEREO-B corotation forecasts are seen to be positively correlated, with a gradient of $1.593~\mathrm{km s^{-1} deg^{-1}}$ and correlation coefficient of $0.301$. This is somewhat expected (\citep{owens2020latRep}) and will be explored in more detail later in the study. (In particular, we note the possibility for aliasing between latitude and longitude separations, as well as with the solar cycle.) Furthermore, the p-value of $0.012$ indicates that this correlation is significant at the $95\%$ confidence level. As the latitude difference between STEREO-B and ACE becomes greater, the STEREO-B observations are sampling solar wind that emanate from different regions of the Sun (e.g., spacecraft could be sampling solar wind from different coronal holes). On the other hand, the posterior forecast does not display this correlation, the p-value indicating that the  correlation found is not significant and the line of best-fit is a flat line (gradient of posterior line of best fit is $0.138~\mathrm{km s^{-1} deg^{-1}}$). This means that when STEREO-B and ACE are at different latitudes, the data assimilation is able to remove this systematic error through combination with the other observation sources (ACE and STEREO-A). However, this does also work against the posterior forecast, as it does not necessarily lead to improvements over the STEREO-B corotation forecast when the latitude differences are small. This could be due to STEREO-A being at a larger latitude offset from ACE, `pulling' the posterior forecast away from the ACE data.

In the next sections, we shall investigate how the data assimilation performs in different solar-cycle regimes and where DA can give the greatest improvements when compared to the STEREO-B forecast.

\subsection{Forecasting the solar-wind speed during different solar-cycle phases}

\begin{figure}
    \centering
    \includegraphics[width=\textwidth]{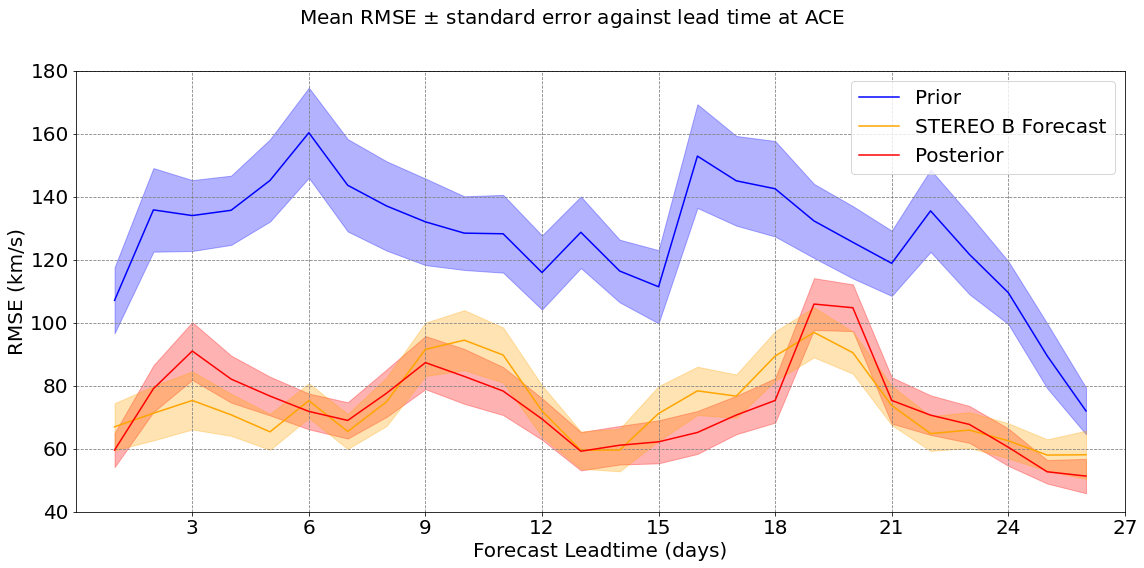}
    \caption{The mean  Root Mean Squared Errors (RMSEs) of the solar-wind speed forecasts plotted against lead-time during solar-minimum ($sC < 0.2$ or $sC \geq 0.7$) at ACE. The blue lines indicate the prior (no DA) forecast RMSEs, the red lines indicate the posterior (including DA) forecast RMSEs and the orange line gives the STEREO-B corotation forecast RMSEs at ACE. The shaded regions indicate one standard error around the mean.}
    \label{fig:meanRMSE27dSolMin}
\end{figure}

\begin{figure}
    \centering
    \includegraphics[width=\textwidth]{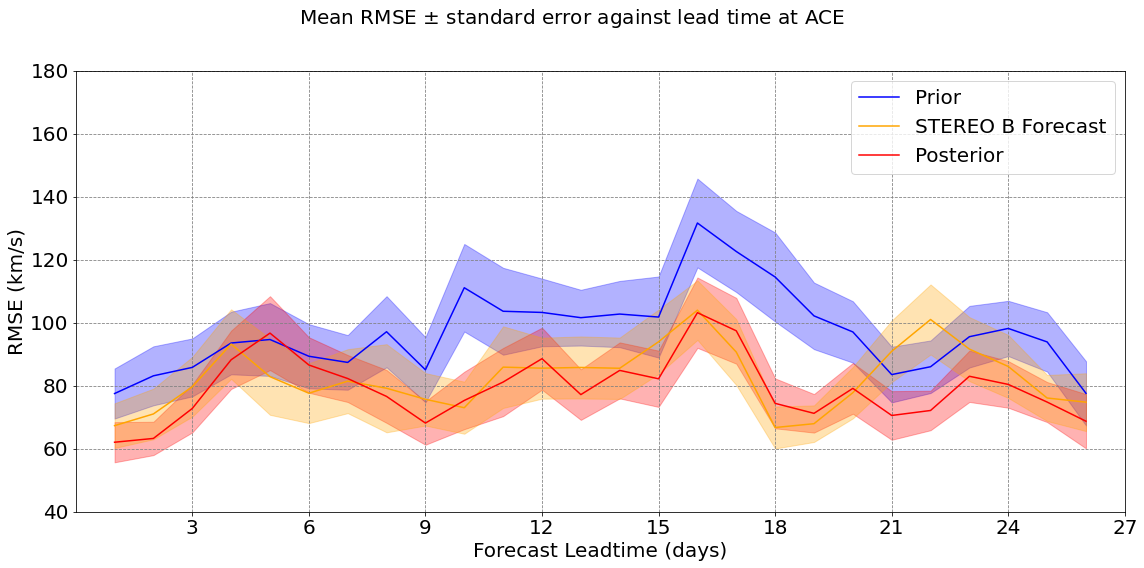}
    \caption{The mean  Root Mean Squared Errors (RMSEs) of the solar-wind speed forecasts plotted against lead-time during solar-maximum ($0.2 \leq sC <0.7$) at ACE. The blue lines indicate the prior (no DA) forecast RMSEs, the solid red lines indicate the posterior (including DA) forecast RMSEs and the solid orange line gives the STEREO-B corotation forecast RMSEs at ACE. The shaded regions span one standard error about the mean.}
    \label{fig:meanRMSE27dSolMax}
\end{figure}

\begin{figure}
	\begin{center}
        \includegraphics[width=\textwidth]{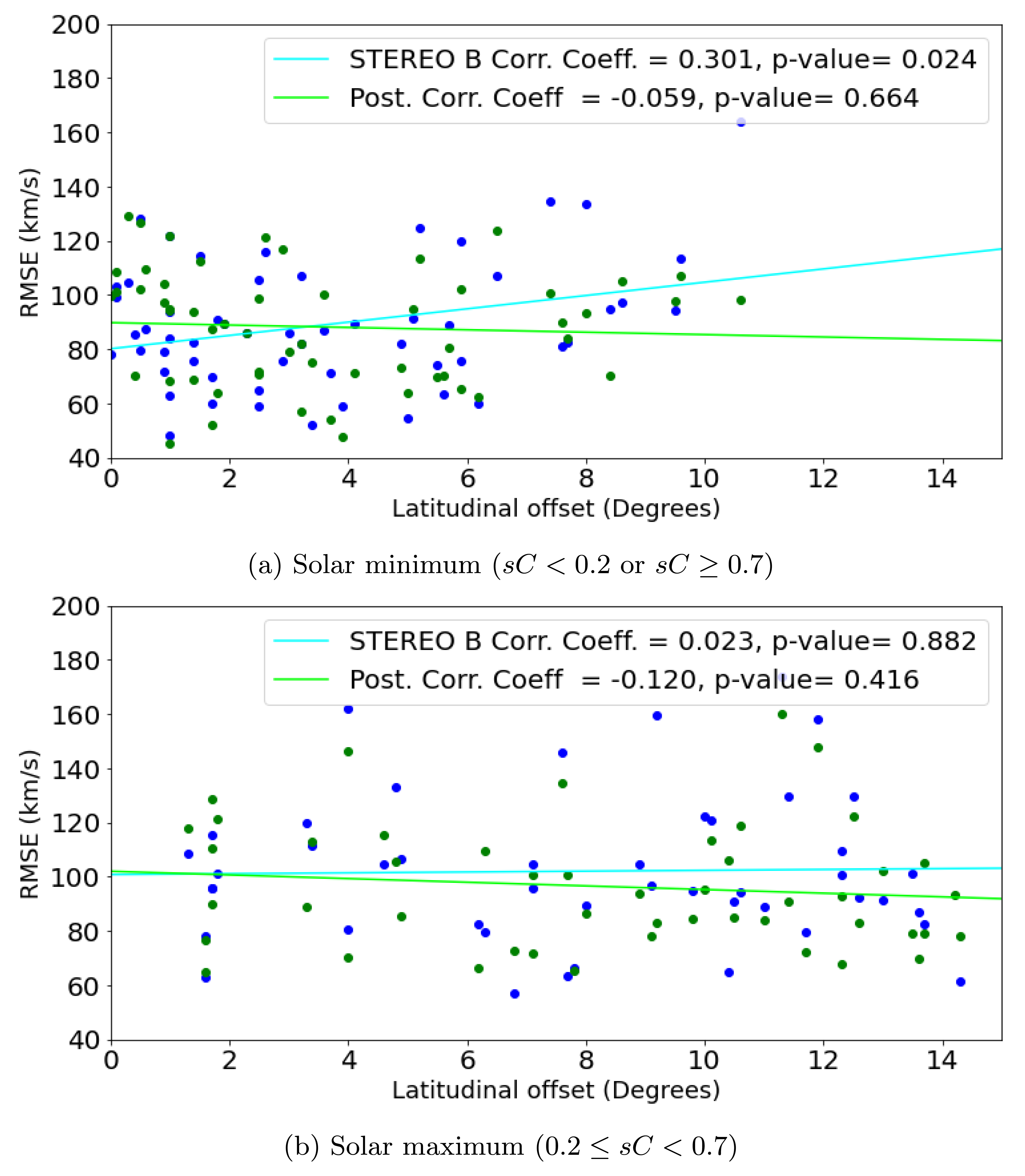}
		\caption{The 27-day Root Mean Squared Errors (RMSEs) of the solar-wind speed forecasts plotted against the absolute difference between STEREO-B and ACE's latitude during solar-minimum (top panel) and solar-maximum (bottom panel). The blue points correspond to the STEREO-B corotation forecast and the green points relate to the posterior forecast. The blue and green lines represent the line of best-fit for the corotation and posterior forecasts, respectively. The legend gives the Pearson's product-moment correlation coefficient indicating the strength of the correlation between the data and the line of best-fit, along with its associated p-value. A significant correlation between latitude offset and RMSE is only observed for the STEREO-B forecast to ACE during solar-minimum (at the $p=0.024$ level).}
		\label{fig:27dlatsolMinMax}
	\end{center}
\end{figure}

\begin{table}
	\begin{center}
		\begin{tabular}{ | >{\centering\arraybackslash}m{1.9cm} | >{\centering\arraybackslash}m{2.5cm} | >{\centering\arraybackslash}m{2.5cm} | >{\centering\arraybackslash}m{3.6cm} | }
			\hline
			 Observation location & Prior RMSE (km/s) & Posterior RMSE (km/s) & STEREO-B corotation RMSE (km/s) \\ \hline
            STEREO-A & $127.146 \pm 12.270$ & $78.056 \pm 7.039$ & N/A \\ \hline
            STEREO-B & $131.010 \pm 12.712$ & $75.096 \pm 7.106$ & N/A \\ \hline
            ACE & $127.145 \pm 12.209$ & $73.386 \pm 6.737$ & $73.802 \pm 7.122$ \\ \hline\hline
            All avg. & $128.434 \pm 12.397$ & $75.513 \pm 6.961$ & N/A \\
			\hline
		\end{tabular}
	\end{center}
	\caption{Table showing mean Root Mean Squared Errors (RMSEs) $\pm$ their standard errors for the prior, posterior and STEREO-B corotation forecasts at each observation location during solar-minimum.}
	\label{tab:meanFcSolMin}
\end{table}

\begin{table}
	\begin{center}
		\begin{tabular}{ | >{\centering\arraybackslash}m{1.9cm} | >{\centering\arraybackslash}m{2.5cm} | >{\centering\arraybackslash}m{2.5cm} | >{\centering\arraybackslash}m{3.6cm} | }
			\hline
			 Observation location & Prior RMSE (km/s) & Posterior RMSE (km/s) & STEREO-B Corotation RMSE (km/s) \\ \hline
            STEREO-A & $99.057 \pm 10.961$ & $83.287 \pm 8.732$ & N/A \\ \hline
            STEREO-B & $96.992 \pm 11.893$ & $84.806 \pm 9.521$ & N/A \\ \hline
            ACE & $96.967 \pm 10.610$ & $79.291 \pm 8.294$ & $82.541 \pm 9.554$ \\ \hline \hline
            All avg. & $97.672 \pm 11.155$  & $82.461 \pm 8.849$  & N/A \\
			\hline
		\end{tabular}
	\end{center}
	\caption{Table showing mean Root Mean Squared Errors (RMSEs) $\pm$ their standard errors for the prior, posterior and STEREO-B corotation forecasts at each observation location during solar-maximum.}
	\label{tab:meanFcSolMax}
\end{table}

Solar activity is expected to influence the forecast accuracy of both the DA and corotation forecasts. During solar-maximum there is an increased number of transient events, such as coronal mass ejections (CMEs), and a more rapidly evolving solar wind structure. Conversely, solar-minimum results in fewer transients and a more slowly evolving solar wind structure. Solar-wind speed is highly structured in latitude near solar-minimum. The solar cycle (from one solar-minimum to the next) is approximately $11$ years. In this section, we denote the phase of the solar cycle as $s_C \in [0,1]$. $0$ denotes the start of the solar cycle (at solar-minimum), $1$ denotes the end of the solar cycle (at solar-minimum) and $0.5$ is the mid-point of the solar cycle. solar-maximum typically occurs prior to $s_C = 0.5$.

We split forecasts of the solar-wind speed into two groups: those around solar-minimum ($sC \geq 0.7$ or $sC<0.2$) and those forecast around solar-maximum ($0.2 \leq sC <0.7$). From this definition, solar-minimum corresponds to dates between 01 January 2007 (the start of the experiments) to 20 March 2011 and solar-maximum from 20 March 2011 to 02 November 2014 (the end of the experiments). In this way, we can diagnose how the solar-wind forecasts behave during these two regimes of the solar cycle. Unfortunately, however, this will not be a like-for-like comparison as the STEREO-B spacecraft is moving away from Earth at a rate of $22.5^{\circ}\mathrm{/yr}$, therefore the corotation forecasts produced later in the seven-year period (i.e., during the solar-maximum phase) will be of longer lead-time than those produced earlier. This means that we would expect the RMSEs to be larger during the solar-maximum phase than those in solar-minimum, due to the fact that they occur when STEREO-B is further away and hence forecasts have a longer lead-time. This is in addition to the more rapidly evolving solar wind structure and increased number of solar wind transients near solar-maximum.

The seven years that we have STEREO-B observations and have run these solar-wind speed forecasts, from $2007-2014$, coincide with the end of Solar Cycle $23$ (August 1996 - December 2008, $s_C>0.84$)  and the first half of Solar Cycle $24$ (December 2008 - May 2020, $s_C<0.51$). This means that we have far more forecasts occurring around solar-minimum than those around solar-maximum.

Tables \ref{tab:meanFcSolMin} and \ref{tab:meanFcSolMax} show that during the solar-minimum regime, the prior solar-wind speed RMSEs and uncertainties are greater than during the solar-maximum regime, as seen in the previous time series plots. During solar-minimum, the posterior forecast and the STEREO-B corotation forecast show a greater improvement over the prior state and greater reduction in uncertainty.

Figure \ref{fig:meanRMSE27dSolMin} shows that during solar-minimum, the posterior and the STEREO-B corotation forecasts show a very similar RMSE pattern for all lead-times, with a mean RMSE difference of $0.086~\mathrm{km/s}$ over all lead-times. Nonetheless, during solar-minimum, the mean posterior forecast RMSEs are lower than the STEREO-B corotation forecast RMSEs for $17$ considered lead-times out of $27$. During solar-maximum (see Figure \ref{fig:meanRMSE27dSolMax}), however, the mean RMSE of the STEREO-B corotation forecast is higher than the posterior forecast RMSE between $20$ and $23$ days (by an average of $17.393 ~\mathrm{km/s}$) and with greater uncertainty (mean standard deviation is $13.95~\mathrm{km/s}$ during this period). Despite this large difference in mean RMSEs, the mean STEREO-B corotation forecasts are better than the posterior forecasts for $11$ of the $27$ lead-time days. On the whole, Figures \ref{fig:meanRMSE27dSolMin} and \ref{fig:meanRMSE27dSolMax} show that the STEREO-B corotation forecast and the posterior forecast RMSEs are broadly similar during solar-minimum and maximum, exhibiting similar means and uncertainty patterns.

By contrast, Figure \ref{fig:27dlatsolMinMax} shows a very different picture of forecasting skill as a function of latitude offset during solar-minimum and solar-maximum. Figure \ref{fig:27dlatsolMinMax}(a) shows the $27$-day RMSEs of the STEREO-B corotation forecast and posterior forecasts against latitude offset during solar-minimum. It can be seen that there is a significant (at the $95\%$ confidence level) positive correlation between STEREO-B corotation forecast RMSEs and STEREO-B's latitude difference from Earth during solar-minimum (with gradient of $3.078 \mathrm{km s^{-1} deg^{-1}}$). By comparison, during solar-maximum, despite STEREO-B reaching a greater latitude offset from ACE, the STEREO-B corotation forecast RMSE is uncorrelated with the latitude offset of STEREO-B (in fact, the line of best-fit's gradient is $-0.013 \mathrm{km s^{-1} deg^{-1}}$, and is not significantly anti-correlated with latitude). The reason for this almost counter-intuitive result can be explained by looking at the solar-wind structure during solar-minimum and solar-maximum \cite{owens2020latRep}. During solar-minimum, the slow-wind band (and heliospheric current sheet) is closely aligned with the heliographic equator ($\pm 20 ^{\circ}$), this means that for a relatively small change in latitude, large deviations in solar-wind structure are observed. Therefore, during solar-minimum it is more likely that when STEREO-B is more latitudinally offset from ACE, it will also be more offset with respect to solar-wind structure. In comparison, during solar-maximum, slow wind extends to high latitude. This means that there are generally smaller gradients of solar-wind structure with latitude, hence STEREO-B and ACE are more likely to be observing the same solar-wind structures despite their latitudinal offset. For more details on this phenomena, see \citep{owens2020latRep}.

Reassuringly, the data assimilation is able to compensate for this latitudinal correlation and produce a more consistent forecast for all latitudinal offsets between STEREO-B and ACE. It can be seen that during solar-minimum, when this effect is greatest, the posterior forecast's line of best fit is almost flat (its gradient is $0.315 km s^{-1} deg^{-1}$) with very little correlation. As noted before in the plot of latitudinal offset against RMSE for all the data, this does mean that higher RMSE can be observed for the posterior forecasts when STEREO-B's latitude is close to that of Earth. This further illustrates the need to investigate and incorporate the latitudinal representivity into the observation error covariance matrix when performing the data assimilation. During solar-maximum, the posterior is barely correlated with the latitudinal offset, nonetheless, it can be seen that there is a consistent reduction in posterior forecast RMSE when compared to the STEREO-B corotation forecast.

\section{Discussion and Conclusions}

The experiments shown in this study illustrate how data assimilation can be used to produce superior forecasts of near-Earth solar-wind speed up to one solar rotation (27 days) ahead. As solar wind speed generally orders solar wind structures, it is expected to provide a general improvement in space-weather forecasting capability.

Over the seven-year period of this study, we see that the data assimilation scheme (BRaVDA) is able to improve forecasts of the solar-wind speed when compared to the prior ensemble (with no DA), reducing RMSE by an average of $36km/s$ at all spacecraft considered. Additionally, the uncertainties present in the forecast estimates are greatly reduced by assimilating data. The posterior forecast RMSE standard errors decreased by an average of $3.1~\mathrm{km/s}$ compared to the prior forecast RMSE standard deviation. This means that we can have increased confidence in the forecasts produced when using data assimilation. During solar-maximum, this improvement in forecast RMSE is reduced to about $14~\mathrm{km/s}$, but during solar-minimum, this is increased to about $48~\mathrm{km/s}$. This shows that the application of data assimilation, even with a relatively simple solar-wind model, can lead to large improvements in forecasting accuracy. The lack of significant difference in forecast accuracy between solar-minimum and solar-maximum is likely due to competing effects taking precedence. At solar-minimum, the highly structured nature of the solar wind with latitude produces forecast errors for both the corotation and data assimilation approaches. Additionally, the solar-wind speed is generally higher and more variable at solar-minimum than solar-maximum, which also inflates forecast RMSE. Conversely, at solar-maximum, forecast errors are produced by the increased frequency of transient solar-wind structures, which none of the studied forecast approaches incorporate. Also, the STEREO spacecraft are further from Earth during the solar-maximum period studied, which increases forecast error.

The data assimilation forecasts at Earth are of similar accuracy and uncertainty to the STEREO-B corotation forecast accuracy and uncertainty. This is somewhat to be expected as the data assimilation is also based upon the STEREO-B observation and is expected to recreate it at the satellite location. But the DA approach is advantageous for two reasons. Firstly, the data assimilation scheme provides far more information than the STEREO-B corotation forecast. Whilst corotation gives us a point forecast at Earth (or can be corotated to any other point at the same latitude and heliocentric distance), the DA scheme maps observational data throughout the model domain. This means that the DA scheme is able to give us improved estimates of the solar wind throughout the inner heliosphere. The advantage of this is that this will give a more accurate representation of the solar wind between the Sun and Earth, which will provide a more suitable medium for CMEs to be propagated through (e.g., see \citet{owens2020huxt, barnard2020ens}). This will be elaborated upon more in a later study.

Secondly, the data assimilation also provides a more consistent and robust forecast for the solar wind compared to the STEREO-B corotation forecast. DA is able to reduce the representivity errors that are caused by the latitudinal offset between STEREO-B and ACE/Earth. The STEREO-B corotation forecast is sensitive to the latitudinal offset between STEREO-B and Earth, with poorer forecasts occurring when STEREO-B is at greater latitudinal separation. This is restricted to solar-minimum, when there are greatly differing solar-wind regimes at different latitudes. By relying on more than one observation source and accounting for potential errors in the observations, data assimilation is able to compensate for these errors and remove the correlation between forecast error and latitude offset. The DA scheme will also reduce the effect of transient structures (such as coronal mass ejections) seen only at STEREO-B, thus reducing false positive forecasts of high-speed streams. More study is required to determine how to balance the representivity errors contributed from latitudinal offset and the ``age'' of observations for the purposes of forecasting at Earth. Both of these effects are expected to have competing solar-cycle variations.

HUXt is intended to act as a surrogate model for more complex MHD models. This study shows that BRaVDA could be used to define an optimum set of boundary conditions using solar wind observations which can then be used to drive more complex models, such as ENLIL and HelioMAS.

%
%
%
%

%
%
%
%
%
%
%

\acknowledgments
We are grateful to the Space Physics Data Facility and National Space Science Data Center for OMNI. STEREO-A, STEREO-B and ACE data were downloaded from the OMNIWeb portal at \url{https://omniweb.gsfc.nasa.gov/}. MAS model output is available from the Predictive Science Inc. website: (http://www.predsci.com/mhdweb/home.php). Work was part-funded by Science and Technology Facilities Council (STFC) grant number ST/R000921/1, and Natural Environment Research Council (NERC) grant number NE/P016928/1. Jake Witherington was funded by a RAS undergraduate bursary. Harriet Turner is funded through SCENARIO grant number NE/S007261/1. HUXt can be downloaded in the Python programming language from \url{https://doi.org/10.5281/zenodo.4889327}. BRaVDA can be downloaded in the Python programming language from \url{https://doi.org/10.5281/zenodo.4849682}.






%
%
%
%
%
%
%
%
%
%

\addcontentsline{toc}{chapter}{Bibliography}
\bibliographystyle{authordate}

\clearpage

\end{document}